\documentstyle[12pt,rotate]{article}
\begin{document}
\def\beq{\begin{equation}}
\def\eeq{\end{equation}}
\def\bea{\begin{eqnarray}}
\def\eea{\end{eqnarray}}
\def\ve{\vert}

\def\nnb{\nonumber}
\def\ga{\left(}
\def\dr{\right)}
\def\aga{\left\{}
\def\adr{\right\}}
\def\rar{\rightarrow}
\def\nnb{\nonumber}
\def\la{\langle}
\def\ra{\rangle}
\def\ba{\begin{array}}
\def\ea{\end{array}}
\def\qqq{\sqrt}

\title{ {\small { \bf PSEUDOSCALAR NEUTRAL HIGGS BOSON 
PRODUCTION IN POLARIZED $\bf\gamma e~$  COLLISIONS} } }

\author{{\small  M. SAVCI \thanks{e-mail: savci@rorqual.cc.metu.edu.tr}}  \\ 
{\small Physics Department, Middle East Technical University} \\ 
{\small 06531 Ankara, Turkey} }

\date{}

\begin{titlepage}
\maketitle
\thispagestyle{empty}

\begin{abstract}
\baselineskip  0.7cm
We investigate the pseudoscalar neutral Higgs-boson 
${\cal A}^0$ production in the polarized $\gamma e$ 
collisions with two different  center 
of mass energies $\sqrt{s} = 500~GeV$ and $\sqrt{s} = 
1~TeV$. The {\it cross-section} of the process $\gamma e 
\rar e {\cal A}^0$ and the {\it polarization asymmetry} due the 
spin of the initial beams are calculated.

\end{abstract}
\end{titlepage}
\baselineskip  .7cm
\newpage

\setcounter{page}{1}
\section{Introduction}
Despite that the Standard Model $(SM)$ describes very 
successfully all the experimental data within the range 
of energies available today, it has many unsolved 
problems, such as the problem of mass, $CP$ violation and
the number of generations, etc., and the failure of the 
unification of  
electroweak and strong forces. For this reason 
one would expect that a more fundamental theory should 
exist, which describes the three forces within the 
context of a single group. Recent $LEP$ data shows
that this indeed can be achieved in Supersymmetric 
Grand Unified Theories \cite{R1} .

It is well known that the minimal supersymmetric extension 
of the $SM$ ($MSSM$) predicts the existence of many new particles
(supersymmetric partners of the ordinary particles) and the Higgs sector of
the theory contains two Higgs doublets in order to give mass to 
the up and down quarks \cite{R2}. The physical Higgs spectrum of the $MSSM$
is richer than the $SM$ and  
contains two $CP$-even and one $CP$-odd neutral Higgs bosons
and a charged Higgs boson \cite{R3}.    
The $MSSM$ extension of the $SM$ 
leads to clear and distinct experimental signatures especially in the 
Higgs sector. The search for the Higgs sector of the $MSSM$ constitutes one
of the main research fields in the existing and future accelerators to which
theoretical physicists focus their attention.
A lot of theoretical work in the literature has been  devoted to the production 
and decay channels of Higgs particles, in particular the pseudoscalar
neutral Higgs-boson ${\cal A}^0$, in various reactions 
(see for example \cite{R4} and references therein). Here in this work,
we investigate the production of the ${\cal A}^0$ in the
polarized $\gamma e$ beams, namely in the process 
$\gamma e \rar e {\cal A}^0$, at the center of mass energies
$\sqrt{s}=500~GeV$ and $\sqrt{s}=1~TeV$.
Note that production of the ${\cal A}^0$ 
without polarization in $\gamma e$ beams,
is studied in \cite{R5}. 
However, the numerical results presented in that work,
for the cross-sections of the  the unpolarized beams,
are over estimated
and approximately 20 times larger than ours.

The paper is organized as follows. In sect.2 we calculate the
cross-section for ${\cal A}^0$ production. We present our 
numerical results and give a discussion about them in sect.3.

\section{Calculation of the Cross-Section $\bf \gamma e
\rar e {\cal A}^0$ process}

The process $\gamma e \rar e {\cal A}^0$ is described by two 
type of diagrams: box and triangle diagrams. From
quite a number of diagrams, however, the main contributions 
arise from triangle ones with photon exchange. 
This is clear from the fact that, the photon pole diagrams give 
the dominant contributions since $t=0$ is in the physical 
region. Contributions of the triangle diagrams with $Z$-boson 
exchange and box diagrams give negligible contributions
than that of photon exchange diagrams, since these diagrams are 
non-singular at $t=0$. Therefore these contributions are all 
disregarded in this present work. On the other hand since the 
$SUSY$ extension of the $SM$ is considered, the charged chargino
contributions are included.

The amplitude for the pseudoscalar ${\cal A}^0$ production
in the $\gamma e \rar e {\cal A}^0$ process can be written 
as 
\beq
{\cal M} = \frac {4 \alpha^2}{sin\theta_W m_W} \, \bar u(p_2) 
\gamma_\mu u(p_1)\epsilon_\rho \frac {F(t)}{t}\, (p_1-p_2)_\alpha k_{1\beta}
\epsilon_{\mu \rho \alpha \beta}~. \label{eq1}  
\eeq
Here $p_2$ and $p_1$ are the momenta of the final and initial 
electrons, respectively, $k_1$ and $\epsilon$ are the 
photon momentum and polarization, $t=-(p_1-p_2)^2$, and the loop 
factor $F(t)$ is given as,
\bea
F(t)&=&\Big{[} - N_C Q^2_t m^2_t cot(\beta) C_0(t,m_{{\cal A}^0}^2,m_t^2) 
               - N_C Q^2_b m^2_b tan(\beta) C_0(t,m_{{\cal A}^0}^2,m_b^2) \nnb \\
&&             -~m^2_\tau tan(\beta) C_0(t,m_{{\cal A}^0}^2,m^2_\tau)
               + 2 m_W m_1 g_{11} C_0(t,m_{{\cal A}^0}^2,m^2_1) \nnb \\
&&             +~ 2 m_W m_2 g_{22} C_0(t,m_{{\cal A}^0}^2,m^2_2)
              \Big{]}~,
\eea
where, $N_C=3$ is the color factor, $Q_t=2/3$ and $Q_b=-1/3$ are the 
electric charge of the top and bottom quarks, $m_1$ and $m_2$ are 
the chargino masses and $tan(\beta)$ is the ratio of the vacuum
expectation values of the two Higgs Doublets (see for example 
\cite{R3}). The chargino coupling constants $g_{11}$ and $g_{22}$
depend on the elements of the $2 \times 2$ unitary matrices
$U$ and $V$, that appear in diagonalization of the chargino mass 
matrix \cite{R2}:
\\ \\
\beq
\chi =  \left( \begin{array}{cc}
       M & \sqrt{2} m_W sin(\beta) \\
       \sqrt{2} m_W cos(\beta) & \mu
       \end{array}
       \right)~, 
\eeq
\\
where, $M$ and $\mu$ are the symmetry breaking terms. Following \cite{R5}, 
we also consider two different cases: $M \mu >m^2_W sin(2 \beta)$ (case-1)
and $M \mu <m^2_W sin(2 \beta)$ (case-2), for which
\bea
g_{11}&=&~~~\frac{m_W}{m^2_1-m^2_2} \left[m_2+m_1 sin(2 \beta)\right]~, \nnb \\
g_{22}&=&-~\frac{m_W}{m^2_1-m^2_2} \left[m_1+m_2 sin(2 \beta)\right] ~,
\eea
and
\bea
g_{11}&=&~~~\frac{m_W}{m^2_1-m^2_2} \left[-m_2+m_1 sin(2 \beta)\right]~, \nnb \\
g_{22}&=&-~\frac{m_W}{m^2_1-m^2_2} \left[-m_1+m_2 sin(2 \beta)\right] ~,
\eea
respectively. In further analysis we will take $m_1>m_2$.

The scalar function $C_0$ entering in (2) is given as (since in the present
case one of the external particles is a photon),
\beq
C_0\ga t,m^2_1,m^2_2 \dr = \frac{1}{t-m^2_1} \Bigg{[} C\ga \frac {m^2_1}{m^2_2} \dr
- C\ga \frac {t}{m^2_2} \dr \Bigg{]}~,
\eeq
where,
\bea
C(x)&=&\int_{0}^{1} \frac{dy}{y} \,
log \Big{[}1-xy(1-y)-i\epsilon \Big{]} \nnb \\ \nnb \\
&=& \left\{ \begin{array}{ll} 
~~~2 \Big{[} sinh^{-1} \ga \sqrt{-\frac{\displaystyle x}{\displaystyle 4}} \dr 
\Big{]}^2 &~~~~ x\leq 0  \\ \\
-~2 \Big{[} sin^{-1} \ga \sqrt{\frac{\displaystyle x}{\displaystyle 4}} \dr 
\Big{]}^2 &~~~~ 0\leq x \leq 4 \\ \\
~~~2 \Big{[} cosh^{-1} \ga \sqrt{\frac{\displaystyle x}{\displaystyle 4}} \dr \Big{]}^2 - 
\frac{\displaystyle{\pi^2}}{\displaystyle{2}} +
2 i \pi cosh^{-1} \ga \sqrt{\frac{\displaystyle x}{\displaystyle 4}} \dr &~~~~ x \geq 4~.
\end{array} \right.
\eea
Using eq.(1) and performing summation over final electron spin and taking 
into account the polarization of the initial particle for the cross-section,
we get 
\beq
\frac{d\sigma}{d(-t)} = \frac{1}{64 \pi s^2} \ve{\cal M} \ve^2~,
\eeq
where,
\beq
\ve{\cal M}\ve^2=\frac{\alpha^4}{sin^2 \theta_W m_W^2} \frac{\ve F(t)\ve^2}{(-t)}
\Big{[} s^2+u^2+\xi_2 \lambda (s+u)^2 \Big{]}~.
\eeq
Here we take $s=-(p_1+k_1)^2$, $u=-(p_2-k_1)^2$, and $\lambda$ and $\xi_2$ 
are the longitudinal polarization of the electron and circular polarization of
the photon, respectively.
From eq.(8) it follows that the expression for the differential cross-section
contains the $\frac{\displaystyle 1}{\displaystyle t}$ term and therefore in 
order to obtain the total 
cross-section for the subprocess, one  must introduce a convenient cut-off 
parameter in the lower bound of the integration, namely, either the  
the mass of the electron or the cut-off parameter for the angle. By 
imposing the angular cut-off, i.e., 
$\eta=sin^2\ga \frac{\displaystyle{\theta_{min}}}{\displaystyle 2} \dr
\simeq 10^{-5}$
the expression for the total cross-section takes the following form (see also
\cite{R5}):
\beq
\sigma_{\gamma e}(\hat{s})=
\frac {\alpha^4}{64 \pi sin^2\theta_W m_W^2} \int_{\eta \ga \hat{s} - m_{{\cal A}^0}^2 \dr}
^{\ga \hat{s} - m_{{\cal A}^0}^2 \dr} \frac{dy}{y} \Big{[} \hat{s}^2+u^2+\xi_2 
\lambda (\hat{s}+u)^2 \Big{]} \ve F(y) \ve^2~, 
\eeq
where, $\hat{s}=xs$ and $x$ is the momentum fraction of the electron carried out
by the photon. Explicit calculations show that, the contribution of the $Z^0$ 
exchange diagram to the total cross-section is negligible  than 
that of the photon exchange diagram. 

It is well known that the high energy $e^+e^-$ colliders can be converted 
into high energy $\gamma e$ colliders with the help of the back-scattered laser
beams \cite{R6}. So, taking into consideration the distribution function
of the back-scattered photons, the total cross-section for the 
$\gamma e \rar e {\cal A}^0$ process can be calculated by
\beq
\sigma=\int_{\frac{m^2_A}{s}}^{0.83} dx~G_\gamma(x) \sigma(sx)~,
\eeq
where, $\sigma(sx)$ is given by eq.(10), the explicit form of the distribution 
function is \cite{R6}
\bea
G_{\gamma}(x)&=& \frac{D_1}{D_2}~,\nnb
\eea
where,
\bea
D_1&=& 1 - x + \frac{1}{1-x} - \frac{4 x}{\xi \ga 1 - x \dr} +
\frac{4 x^2}{\xi^2 \ga 1 - x \dr^2}~, \nnb \\
D_2&=& \ga 1 - \frac{4}{\xi} -\frac{8}{\xi^2} \dr log \ga 1 +\xi \dr +
\frac{1}{2} + \frac{8}{\xi} - \frac{1}{2 \ga 1+\xi \dr^2}~.
\eea
In our numerical calculations we take $\xi=4.83$.

\section{Numerical Analysis}
We present our numerical results for the cross-sections for various values of 
$tan(\beta)$ in a series of graphs. The dependence of the cross-section on 
$m_{{\cal A}^0}$ at different values of the circular polarization $\xi_2$ of 
the photon is presented in Fig.1 and Fig.2 for case-1 at $\sqrt{s}=500~GeV$
and $\sqrt{s}=1~TeV$, respectively. 
A similar graphical analysis is illustrated for case-2 in Fig.3 and Fig.4. 
In the numerical calculations, for the largest 
chargino mass we take $m_1=250~GeV$ and for $m_2$ we take the largest value
consistent with the restriction $m_1-m_2 \geq m_W \sqrt{2 \left[ 1 \pm 
sin \ga 2 \beta \dr \right]}$, where upper and lower signs correspond to 
case-1 and case-2, respectively. Such restrictions are needed in order to 
ensure that $M$ and $\mu$ are real. The choice we made for $m_1$ and $m_2$  
in this article, is within the range of the chargino masses that are found in the 
$SUSY$ theories. Further, we checked for the contribution coming from $t$, $b$,
$\tau$ and charginos separately. Our calculations show that, for larger values 
of $tan(\beta)$ the contributions arising from $\tau$ lepton and charginos
become important.

Our starting point is the analysis of case-2. For $tan(\beta)=1$, the 
cross-section decreases with increasing values of $m_{{\cal A}^0}$ and
around $m_{{\cal A}^0} \simeq m_t$, it reaches a deep 
minimum value, which from that point on it  starts increasing and 
attains its maximum value at $m_{{\cal A}^0} \simeq 2 m_t$, and
then decreases smoothly again. For larger values of $tan(\beta)$
(i.e., for $tan(\beta)=5,~20$ and $50$), the behavior of the
cross-section is absolutely different and it gets its first
maximum value around $m_{{\cal A}^0} \simeq m_t$, contrary to the 
previous case. The second maximum point  
is again around $m_{{\cal A}^0} \simeq 2 m_t$.
Note that, at $\sqrt{s}=1~TeV$, around the point 
$m_{{\cal A}^0} \simeq 500~GeV$, a new maximum value for the
cross-section shows itself, which is totally absent in the 
$\sqrt{s}=500~GeV$ case. It is interesting to observe that,
the locations of the maximum values of the cross-section
are all independent of $tan(\beta)$. Similar situation holds 
for all other values $\xi_2$ of the circular polarization of the
photon. The different behavior of the cross-section for the small and
large values of $tan(\beta)$ is due to the increase in the  contributions arising 
from the $\tau$-lepton and  charginos. 

Similar arguments hold for case-1 as well, but with the following 
 differences: For almost all values of $tan(\beta)$ 
the cross-section has its maximum value located around the same point 
$m_{{\cal A}^0} \simeq m_t$, while for $tan(\beta)=20$ and 
$tan(\beta)=50$ it has a minimum around the point
$m_{{\cal A}^0} \simeq 150~GeV$.     
  
For all cases the maximum value of the cross-section is about 
$\sim 0.1 fb$. If we assume that the integral luminosity at 
$\sqrt{s}=500~GeV$ is $10~fb^{-1}$ and
at $\sqrt{s}=1~TeV$ is $60~fb^{-1}$ \cite{R6}, we have at most 2 events 
for $\sqrt{s}=500~GeV$ and 12 events for $\sqrt{s}=1~TeV$.

From the analysis of all graphs, we can deduce the following result: For the 
detection of ${\cal A}^0$, the $tan(\beta)=1$ case is more preferable. 
At $\sqrt{s}=500~GeV$ , when $g_{11}=-g_{22}$ ($g_{11}=g_{22}$), it seems
that
it is possible to detect ${\cal A}^0$ with a mass around 
$m_{{\cal A}^0} \simeq m_t$ ($m_{{\cal A}^0} \simeq 2 m_t$). Similar 
situation is present for $\sqrt{s}=1~TeV$, but with one important 
difference. For the case $g_{11}=-g_{22}$ at $tan(\beta)=1$, it is 
possible to detect ${\cal A}^0$ with a mass around $m_{{\cal A}^0} \simeq m_t$
and $m_{{\cal A}^0} \simeq 2 m_t$. Note that for this value of the
$tan(\beta)$, the cross-section for the  case $g_{11}=-g_{22}$, 
is larger than that of $g_{11}=g_{22}$.

Few words about the possibility of detecting ${\cal A}^0$ in experiments
should be mentioned.
In order to detect ${\cal A}^0$ in real experiments, we must take the decay
products of ${\cal A}^0$ into account and separate the background processes
from the signal of the ${\cal A}^0$ decay with the same final state
particles.
The main decay
channels of ${\cal A}^0$ are \cite{R4,R7} :
${\cal A}^0 \rar b \bar b$, $c \bar c$, $t \bar t$, $\tau^+ \tau^-$,
$W^{\pm} W^{\mp}$, $Z h$, $Z H$, $g g$, $\gamma \gamma$ and $Z \gamma$.
If $m_{{\cal A}^0} \leq 2 m_t$, in the minimal supersymmetric models the
dominant decay mode of ${\cal A}^0$ is ${\cal A}^0 \rar b \bar b$ (for more detail
see \cite{R4}). This decay mode is dominant even above the top quark threshold
up to $tan\beta \simeq 30$ \cite{R8}. In this case we can also neglect
the chargino contributions since they are smaller than the one arising from
$b \bar b$ mode (see \cite{R4}). Therefore, in our case the main process 
which is responsible for detecting ${\cal A}^0$ is 
$\gamma e \rar e {\cal A}^0 \rar e b \bar b$ with a branching ratio of
$Br({\cal A}^0 \rar b \bar b) \simeq 0.95$ \cite{R4}. It is clear that in this 
case the main background process is the direct $b \bar b$ production
in $\gamma e \rar e b \bar b$ (see \cite{R5}).
A comparison of our results on cross-section and the background process
$\gamma e \rar e b \bar b$ \cite{R5} leads to the conclusion
that at $\sqrt{s} = 500~GeV$ the background process dominates for all
values of $tan\beta$. On the other hand at $\sqrt{s} = 1 TeV$, and 
$tan\beta = 1$ the $b \bar b$ signal from ${\cal A}^0$ can be detected with
a mass $m_{{\cal A}^0} \simeq 2 m_t$.

Finally we would like to discuss the following question:
Can we deduce any extra information about the mass of ${\cal A}^0$, if we 
investigate the polarization asymmetry due to the longitudinal and 
circular polarization of the initial electron and photon, 
respectively? In order to answer this 
question, we presented in Fig.(5) the behavior of polarization asymmetry versus
$\sqrt{s}$ at different values of $m_{{\cal A}^0}$. We clearly observe that when 
$m_{{\cal A}^0} \simeq m_t$ polarization asymmetry approaches to zero
around $\sqrt{s} \simeq 1.7~TeV$, and this value of energy is far from
the maximum available energy range of the present colliders. Moreover, 
it follows from 
this figure that, the dependence of polarization asymmetry on $\sqrt{s}$ at 
different values of $m_{{\cal A}^0}$, is practically the same. Therefore, 
in our opinion, the polarization asymmetry is not a convenient tool for deducing 
extra information about the mass of ${\cal A}^0$. 

Our final conclusion is that,
in search of the pseudo-scalar ${\cal A}^0$ in $\gamma e$ collider, 
with the preferred choice of $tan(\beta)=1$, it is possible 
to detect the pseudoscalar ${\cal A}^0$ with a mass of 
$m_{{\cal A}^0} \simeq 2 m_t$. 

\newpage

\section*{Figure Captions}

{\bf Fig.1.} The dependence of the cross-section of ${\cal A}^0$ 
             production in $\gamma e \rar e {\cal A}^0$ on the mass  
             $m_{{\cal A}^0}$ for various values
             $tan\beta$ and circular polarization $\xi_2$ of the photon 
             at a fixed value of the longitudinal polarization  of the 
             initial electron ($\lambda = 0.9$), for the case
             $M \mu >m^2_W sin(2 \beta)$ at the center of mass energy 
             $\sqrt{s} = 500~GeV$. 
             The mass of the charginos are given in the text. \\ \\
{\bf Fig.2.} The same as in Fig.1, but for $\sqrt{s} = 1~ TeV$. \\ \\
{\bf Fig.3.} The same as in Fig.1 but for the case  
             $M \mu < m^2_W sin(2 \beta)$. \\ \\
{\bf Fig.4.} The same as in Fig.3  but for $\sqrt{s} = 1~ TeV$. \\ \\
{\bf Fig.5.} Polarization asymmetry 
             ${\cal A} = \frac{\displaystyle{\sigma(\xi_2=+1) -
                                             \sigma(\xi_2=-1)}}
                              {\displaystyle{\sigma(\xi_2=+1) +
                                             \sigma(\xi_2=-1)}}$ 
             versus $\sqrt{s}$ for the case 
             $M \mu < m^2_W sin(2 \beta)$, for different values of
             $m_{{\cal A}^0}$ and at $tan\beta = 1$, $\lambda = 0.9$.
             For all other choices of the parameters $tan{\beta}$, $\xi_2$
             and the case $M \mu >m^2_W sin(2 \beta)$, the graphical analysis 
             yields almost identical figures.

\begin{figure}

\vspace{19.0cm}
    \includegraphics{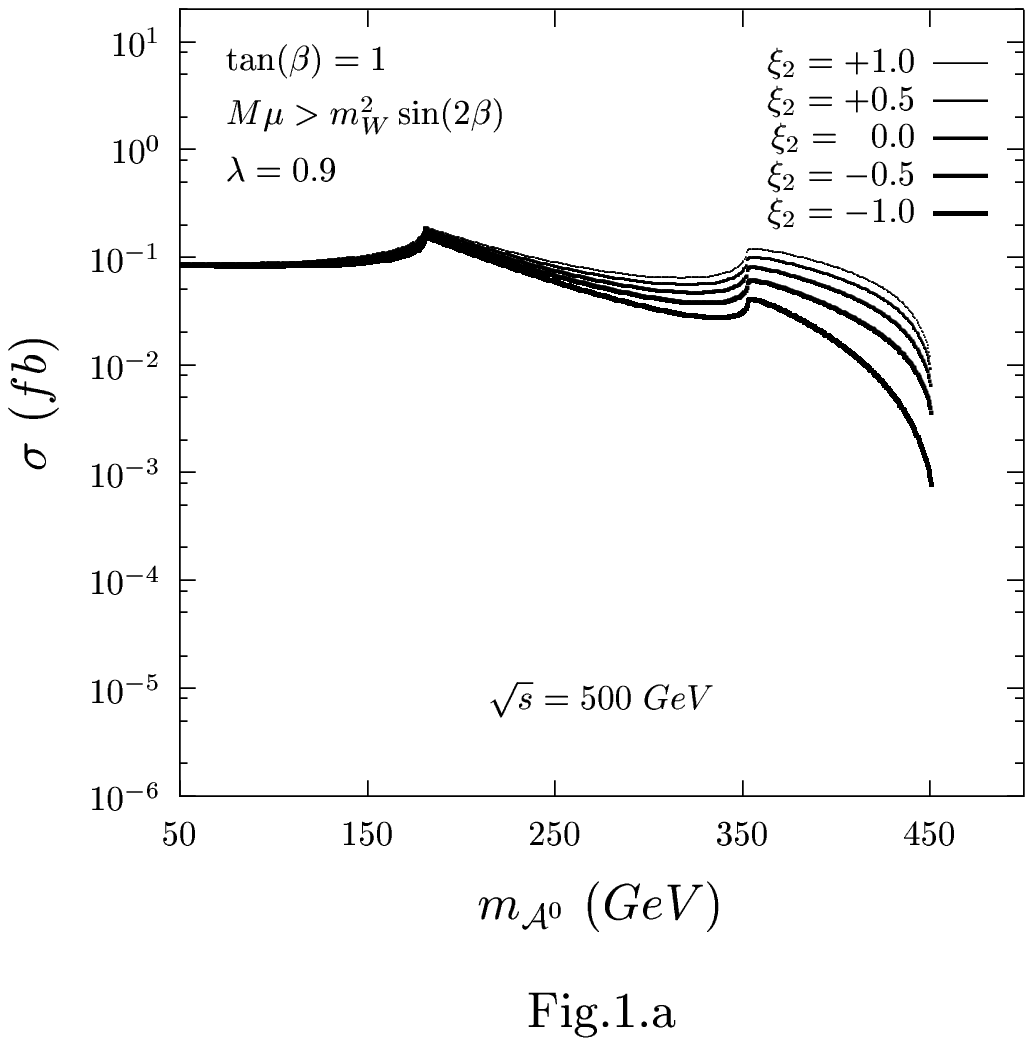}
    \includegraphics{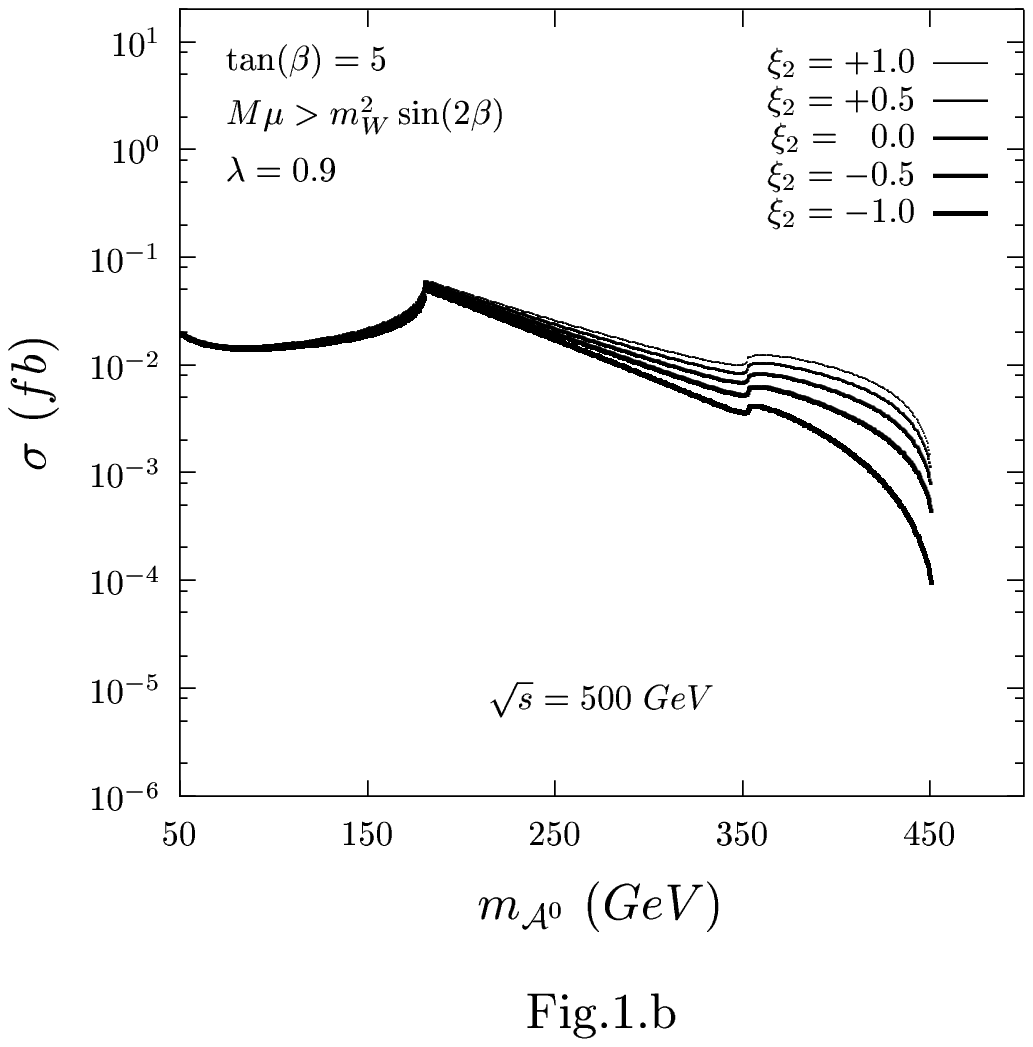}
    \includegraphics{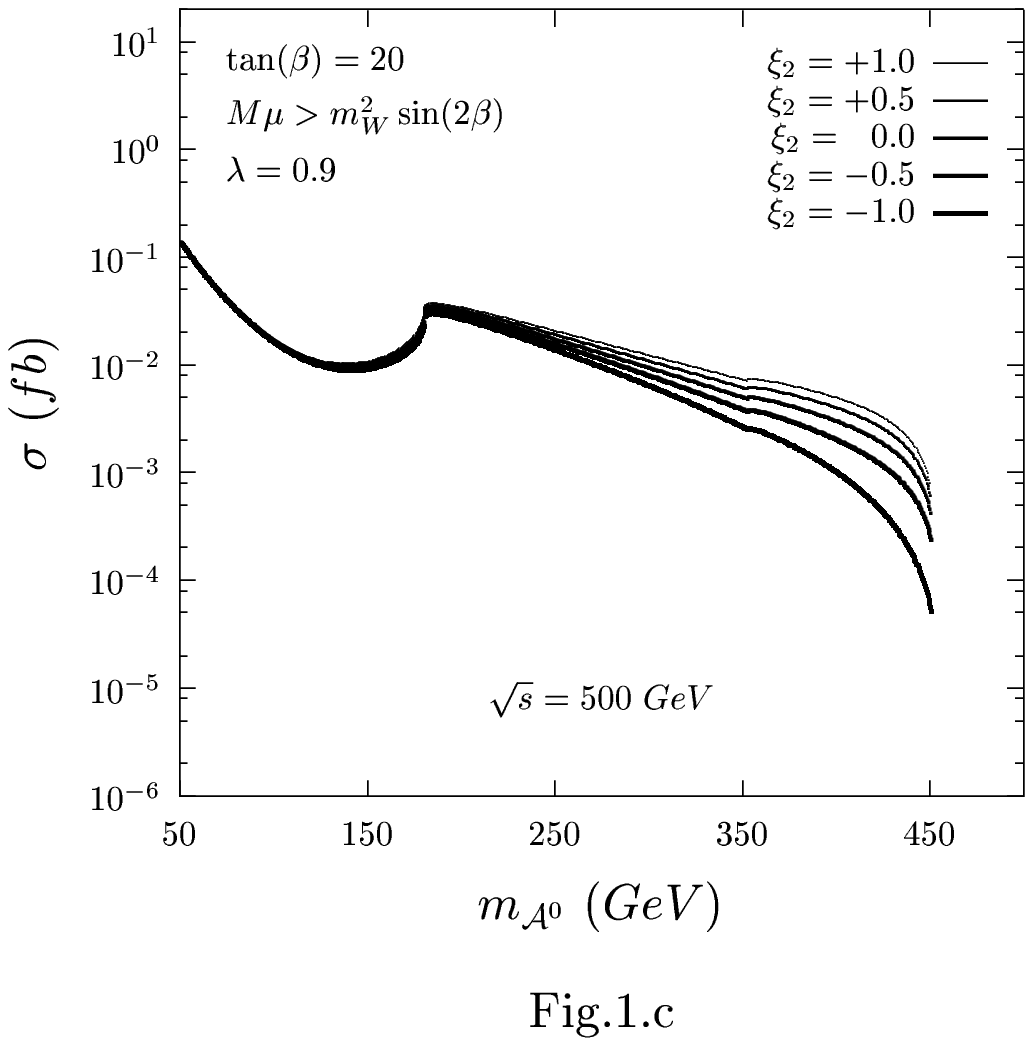}
    \includegraphics{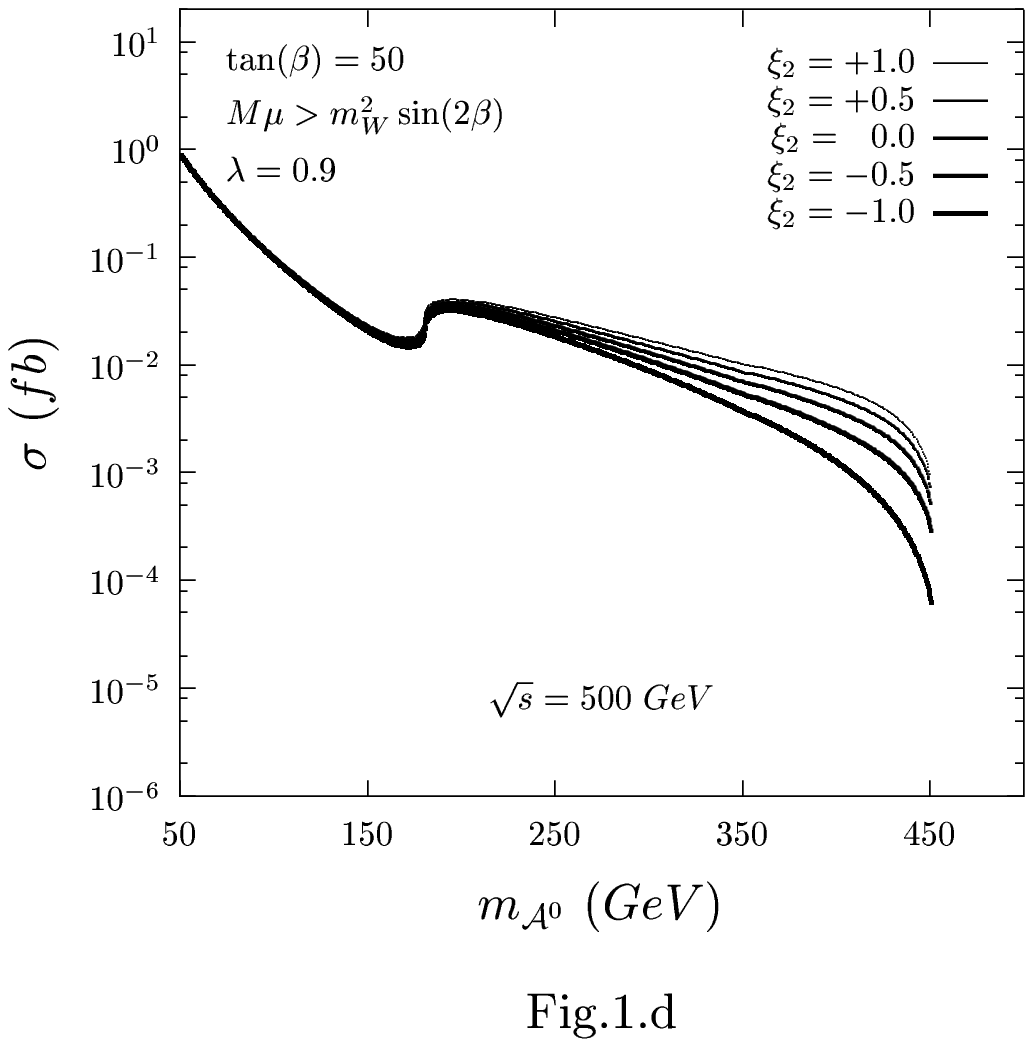}
    \vspace{-2.0cm}

\hspace{4.5cm} Figure 1: Case-1, for $\sqrt{s}=500~GeV$.

\end{figure}

\begin{figure}

\vspace{19.0cm}
    \includegraphics{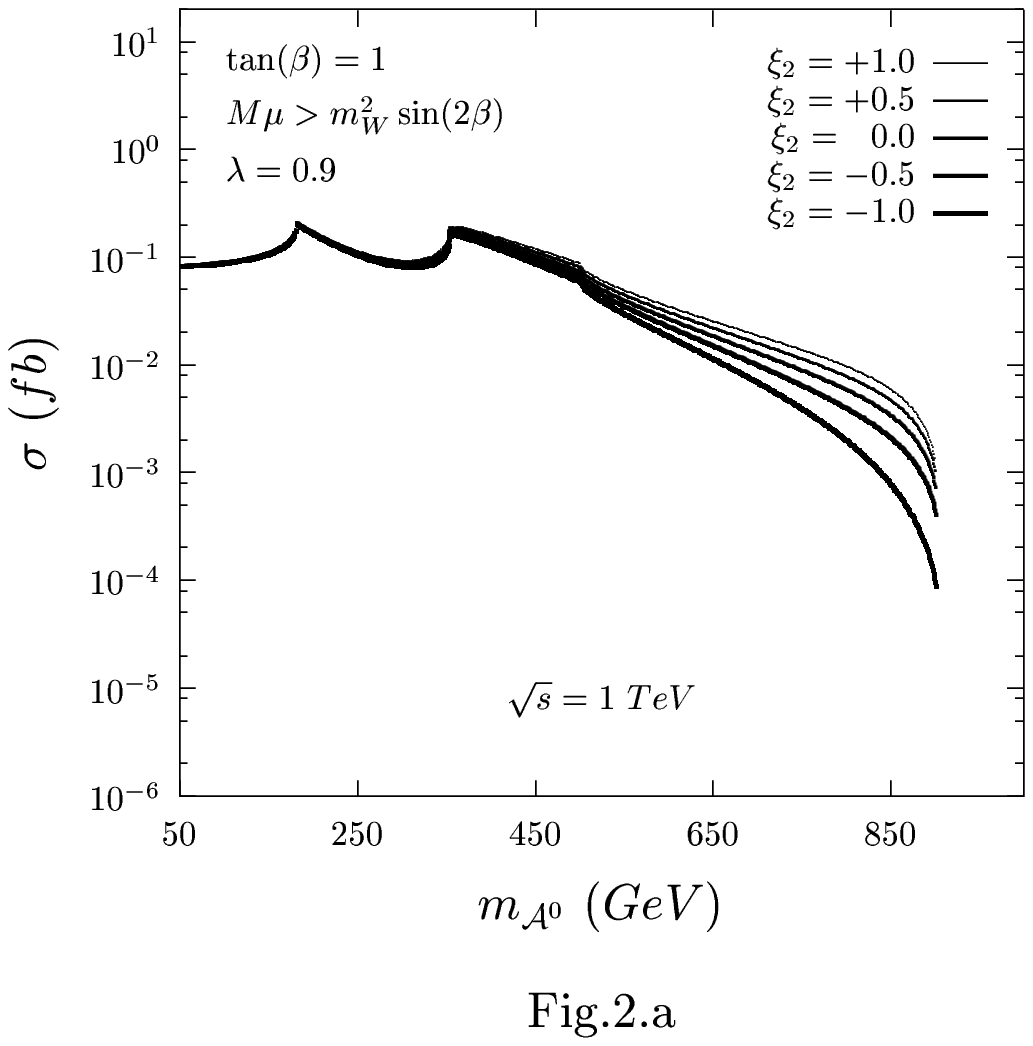}
    \includegraphics{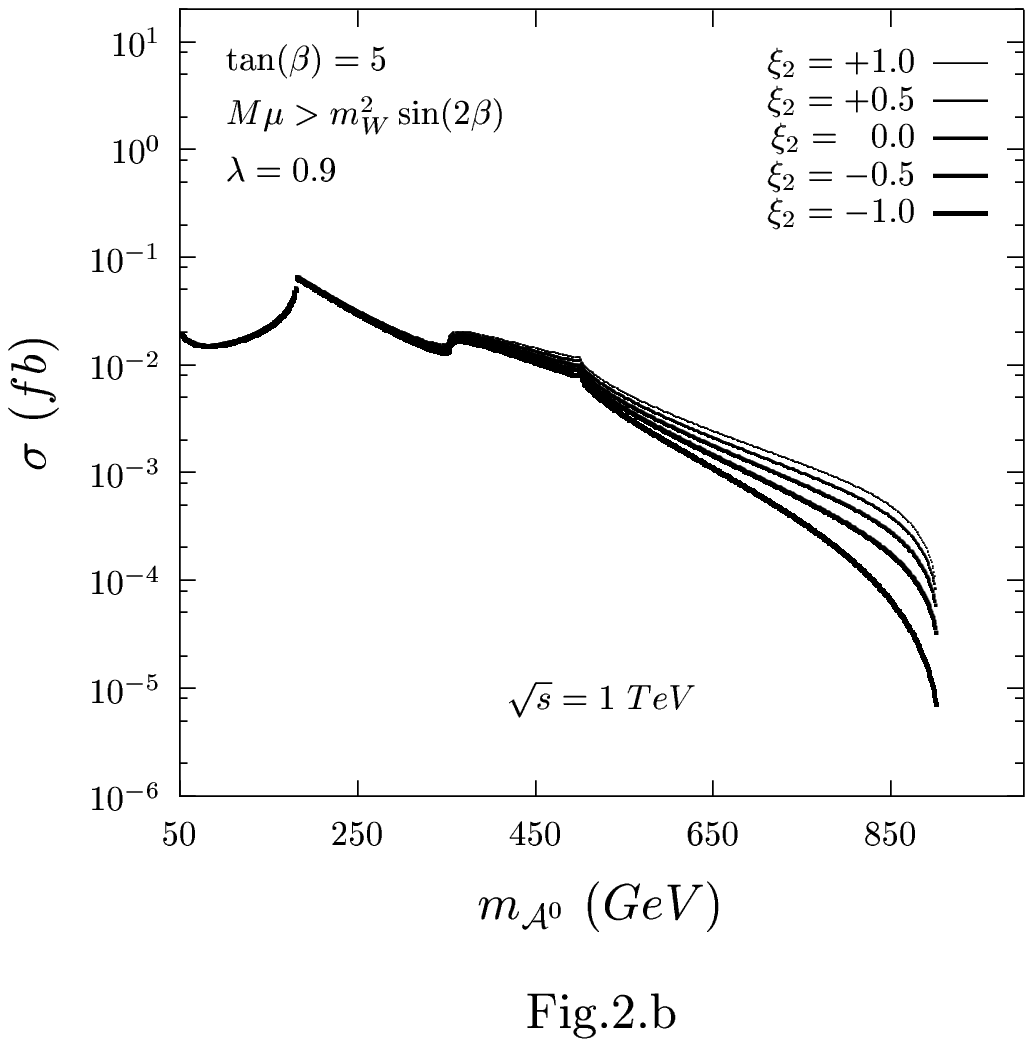}
    \includegraphics{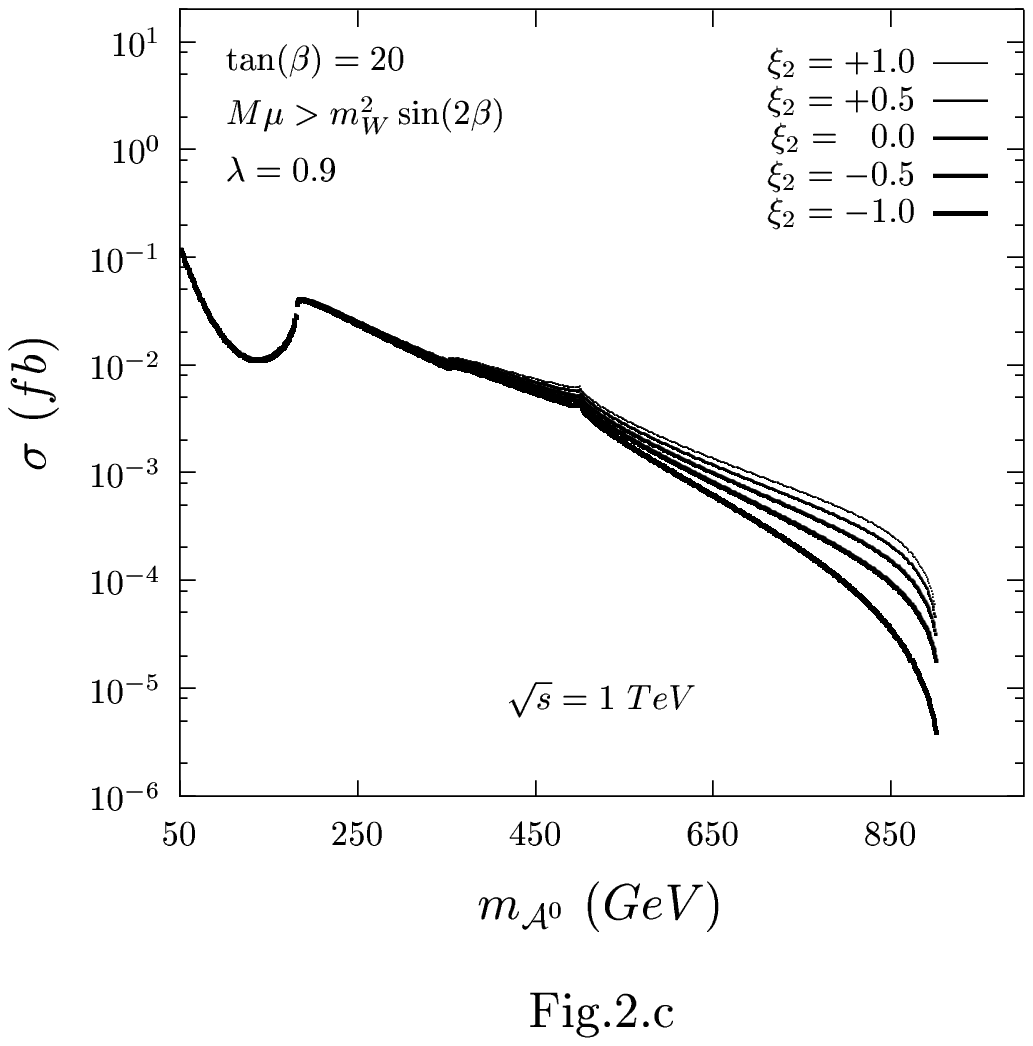}
    \includegraphics{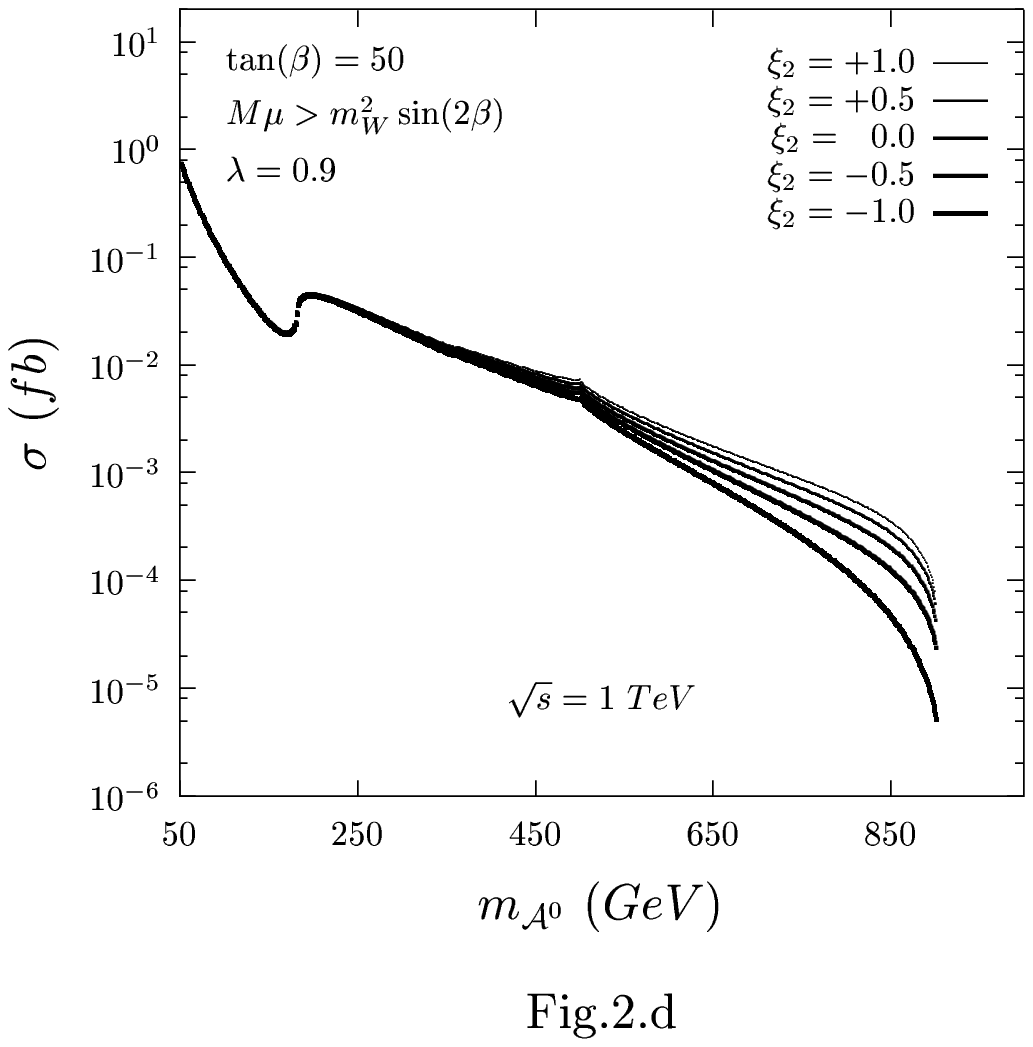}
    \vspace{-2.0cm}

\hspace{4.5cm} Figure 2: Case-1, for $\sqrt{s}=1~TeV$.

\end{figure}

\begin{figure}

\vspace{19.0cm}
    \includegraphics{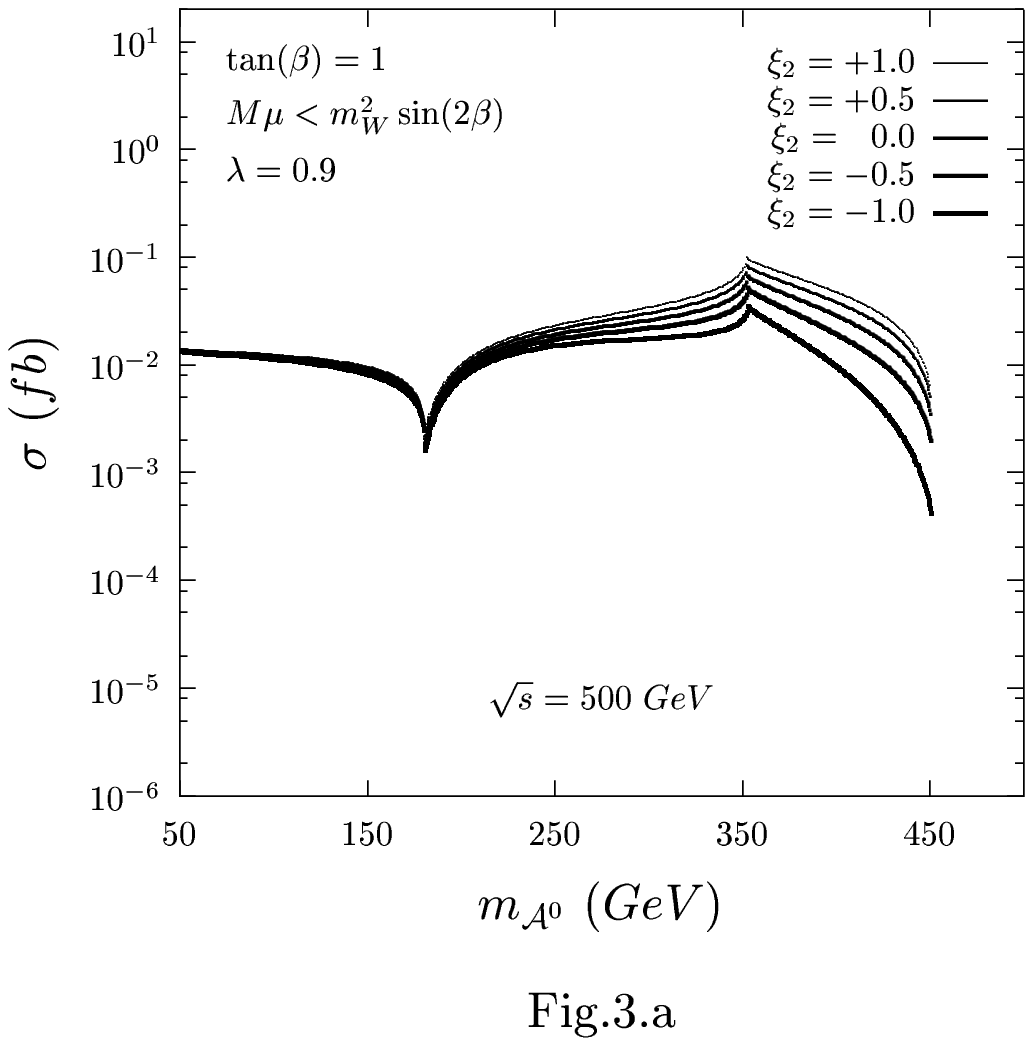}
    \includegraphics{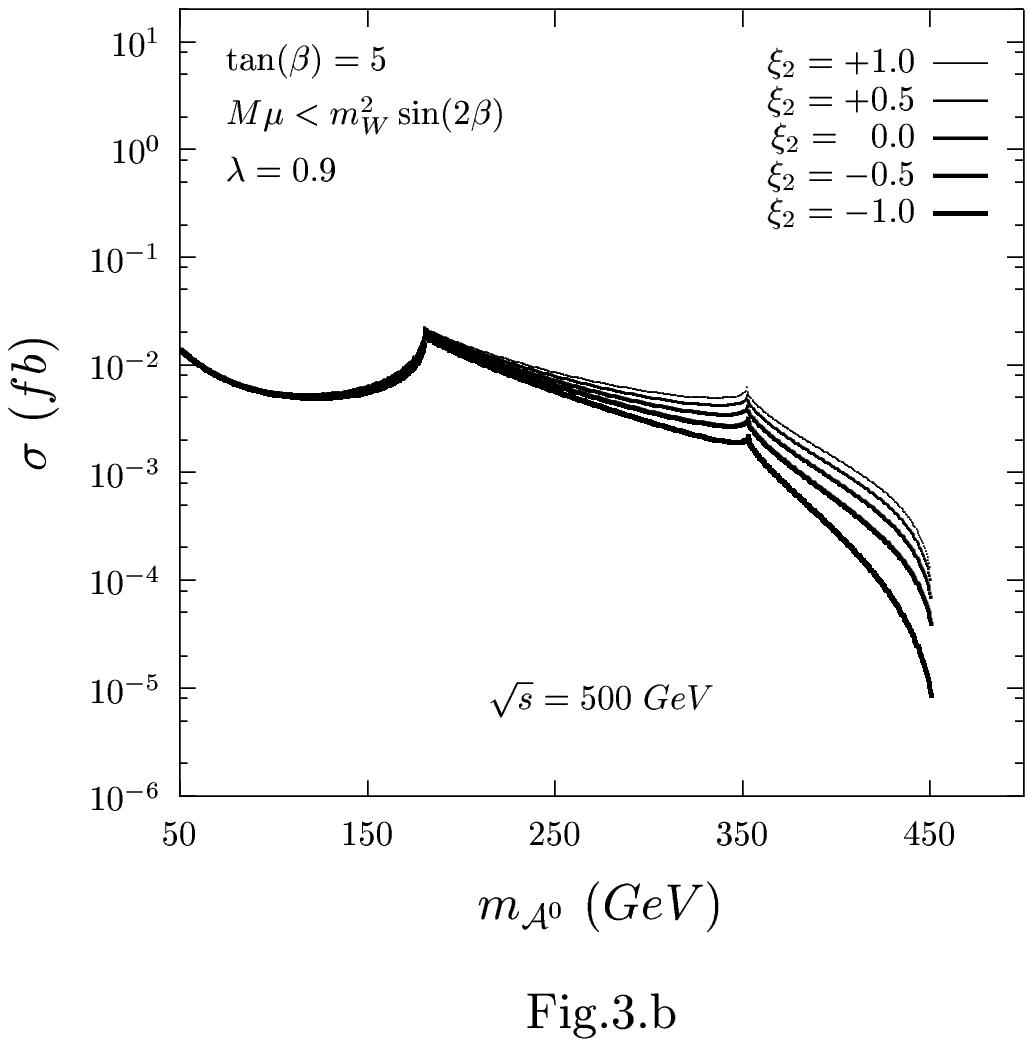}
    \includegraphics{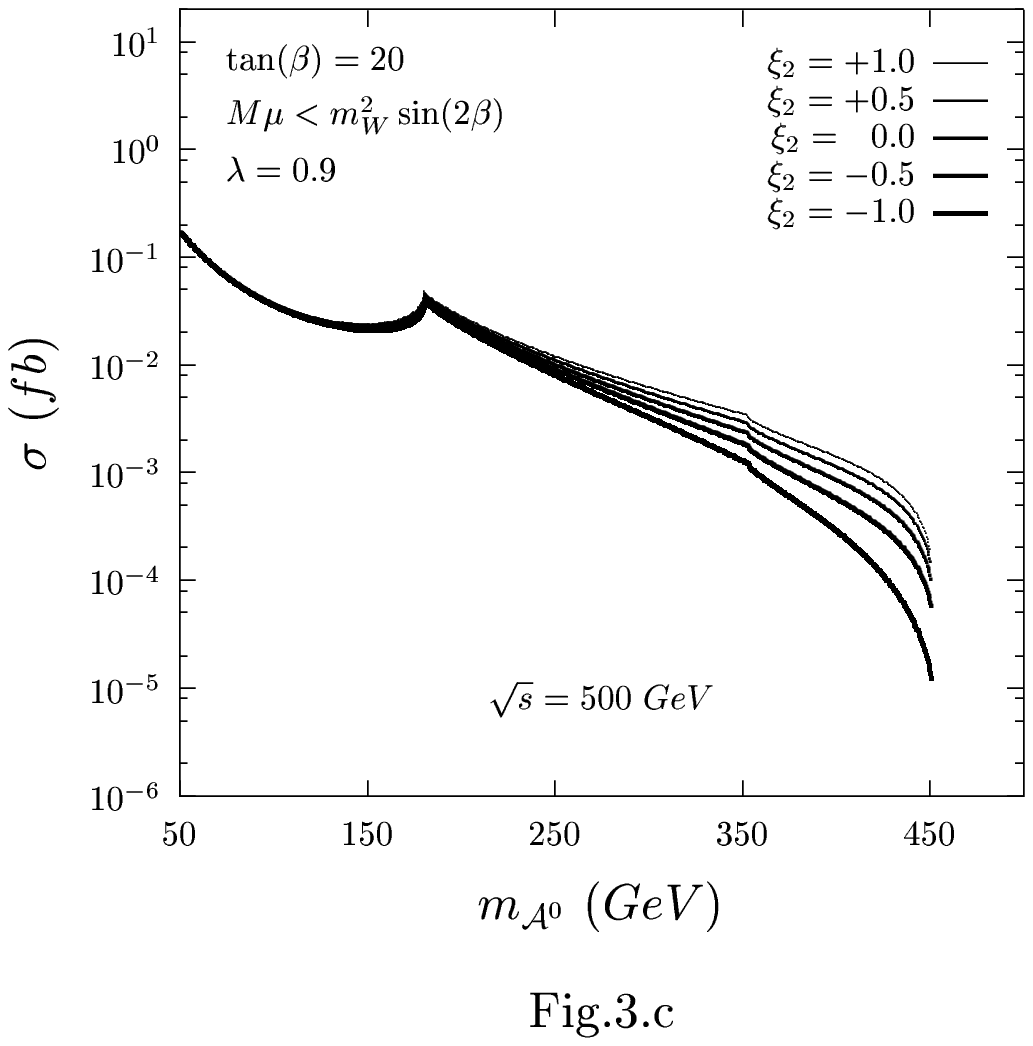}
    \includegraphics{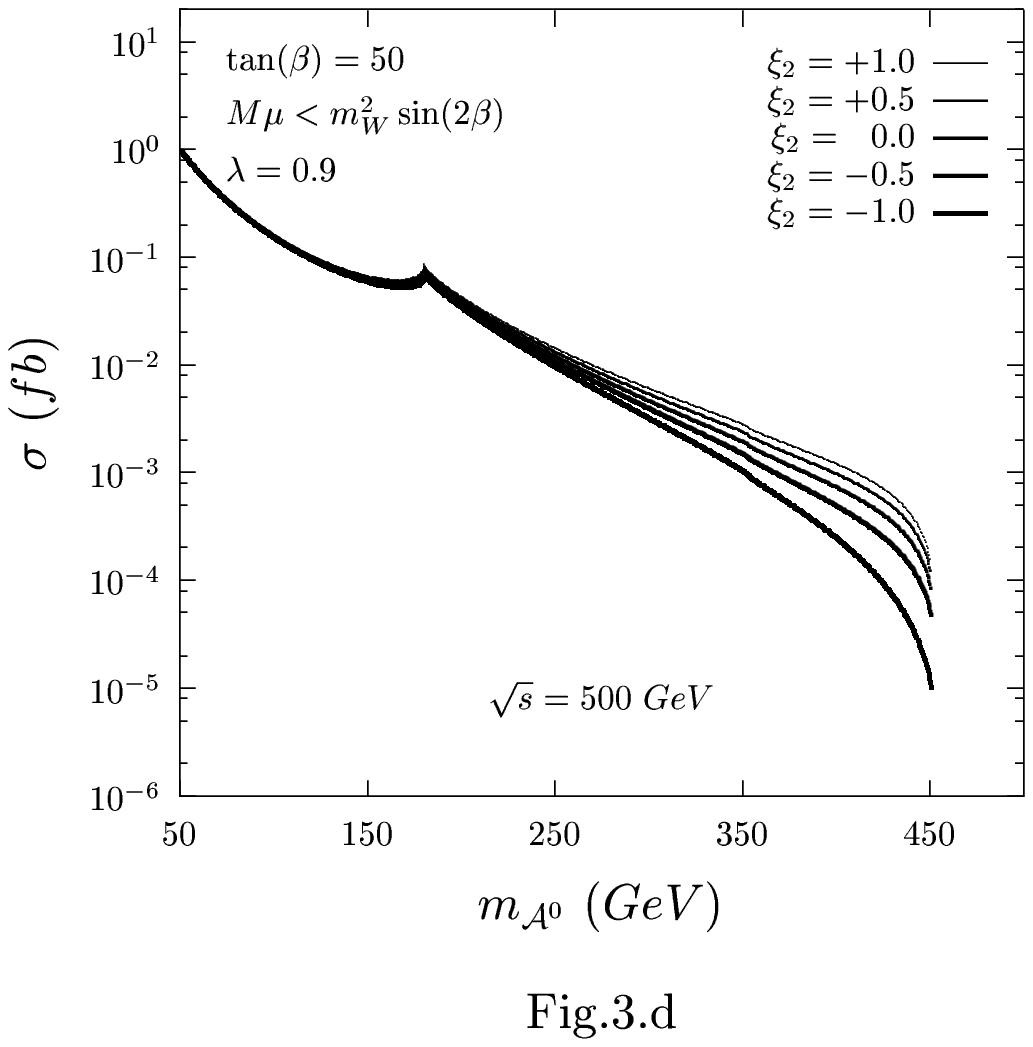}
    \vspace{-2.0cm}
                    
\hspace{4.5cm} Figure 3: Case-2, for $\sqrt{s}=500~GeV$. 
                                                               
\end{figure}

\begin{figure}

\vspace{19.0cm}
    \includegraphics{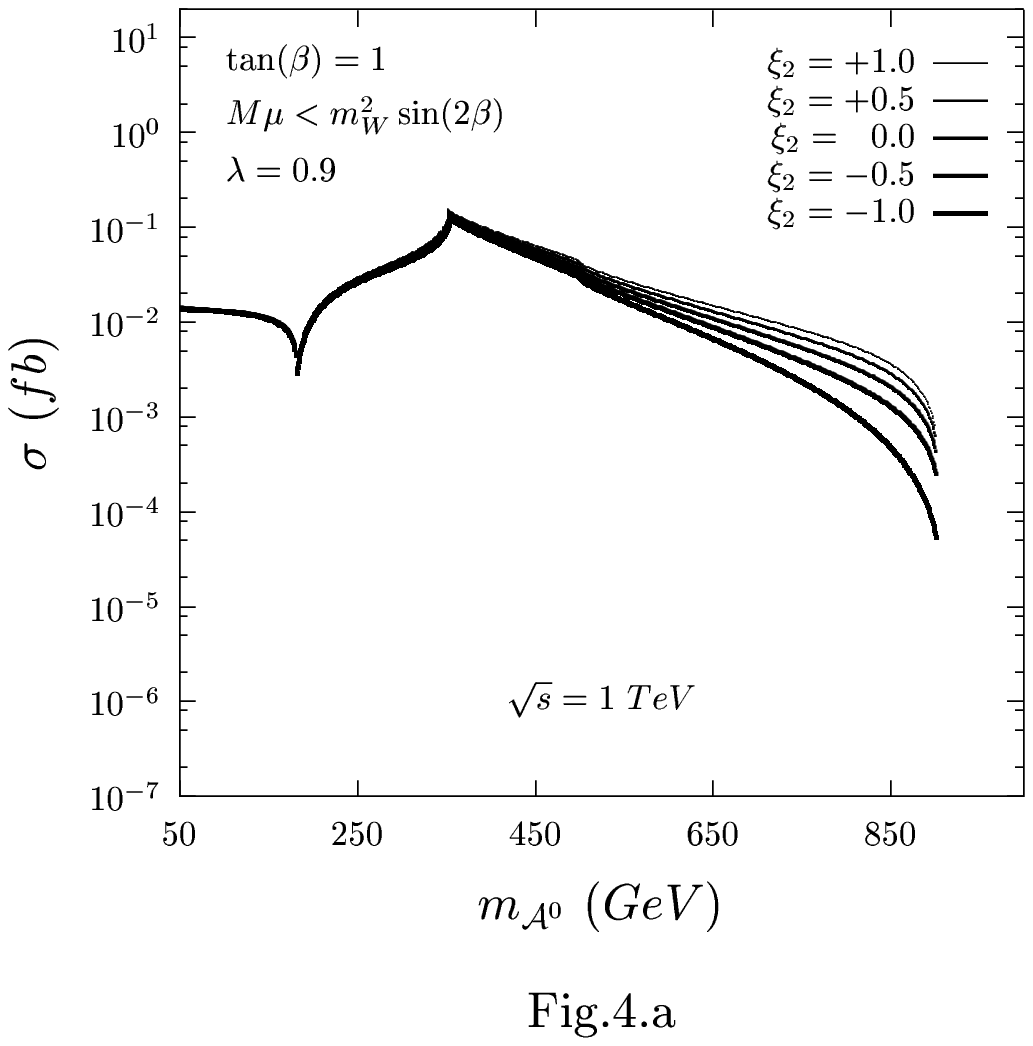}
    \includegraphics{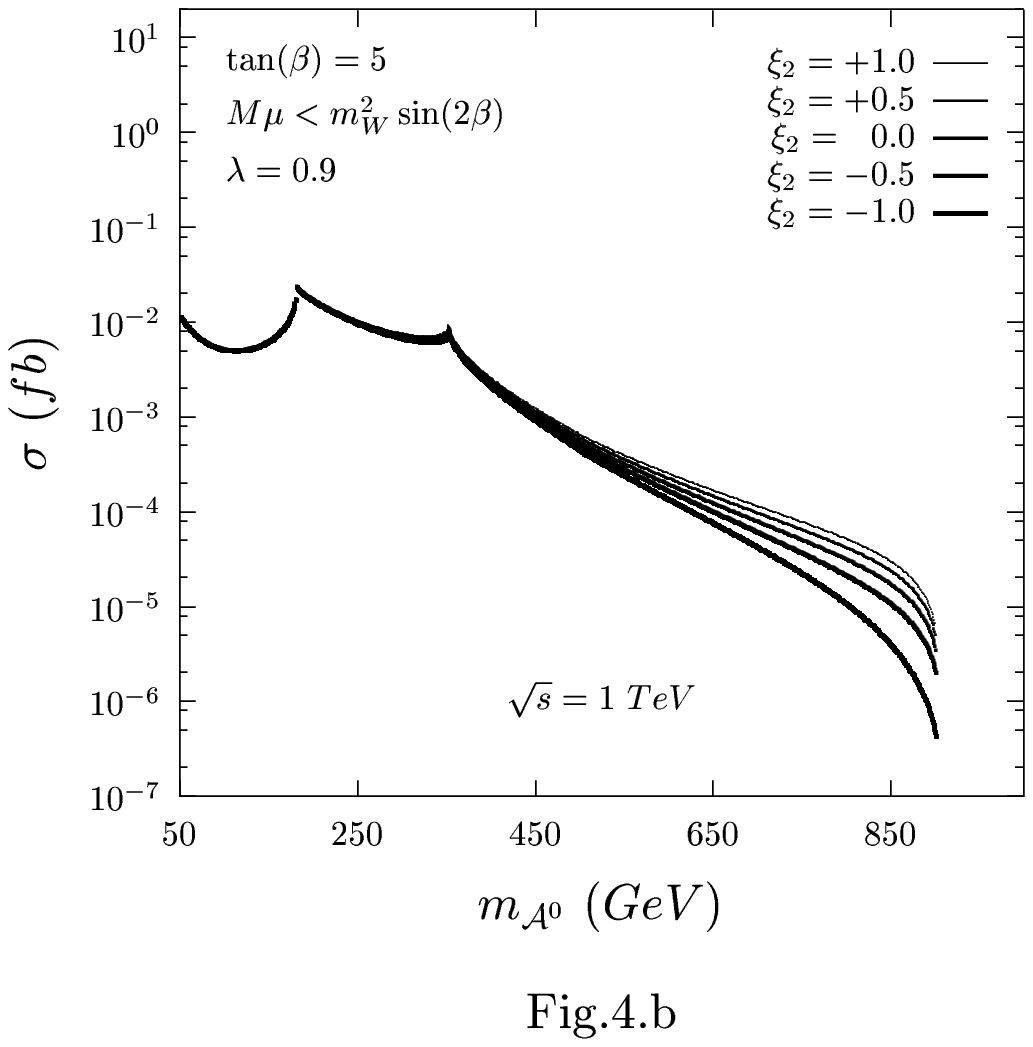}
    \includegraphics{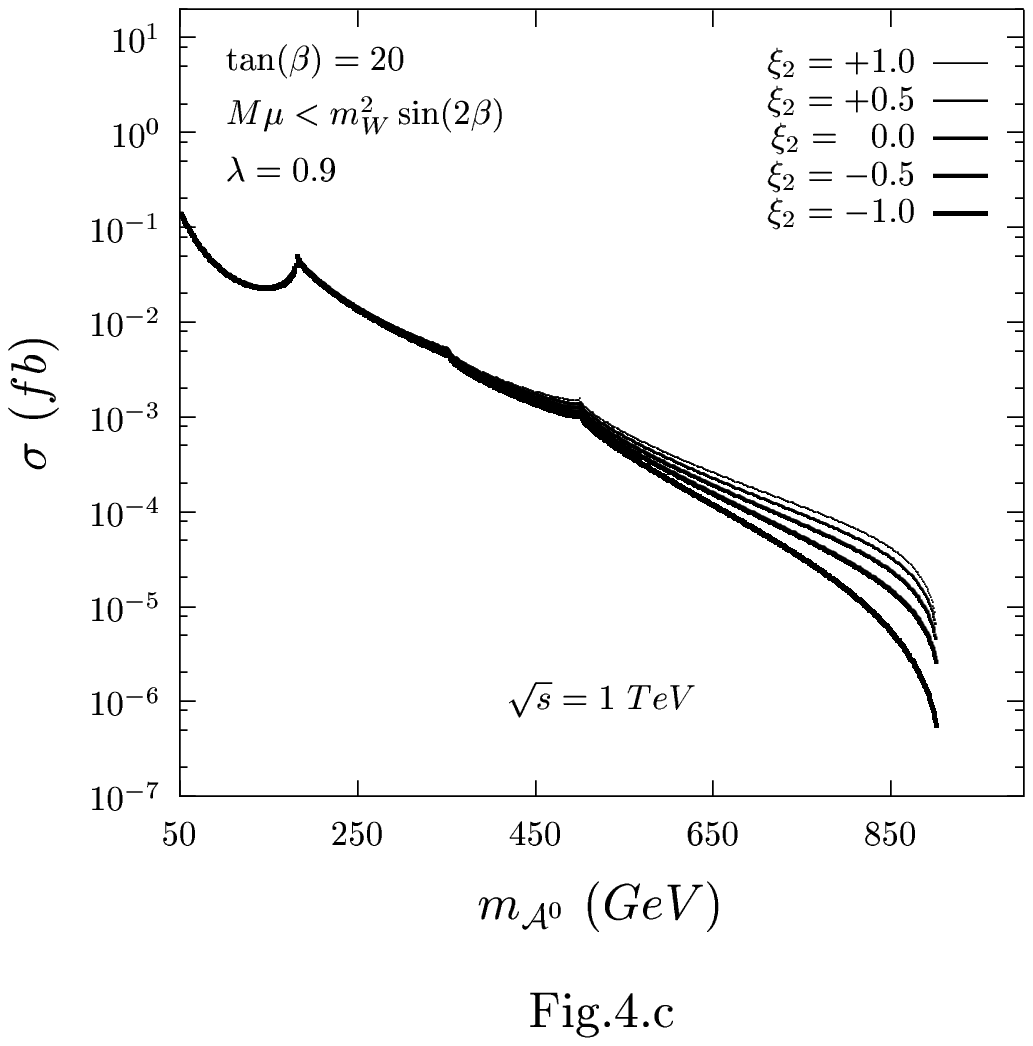}
    \includegraphics{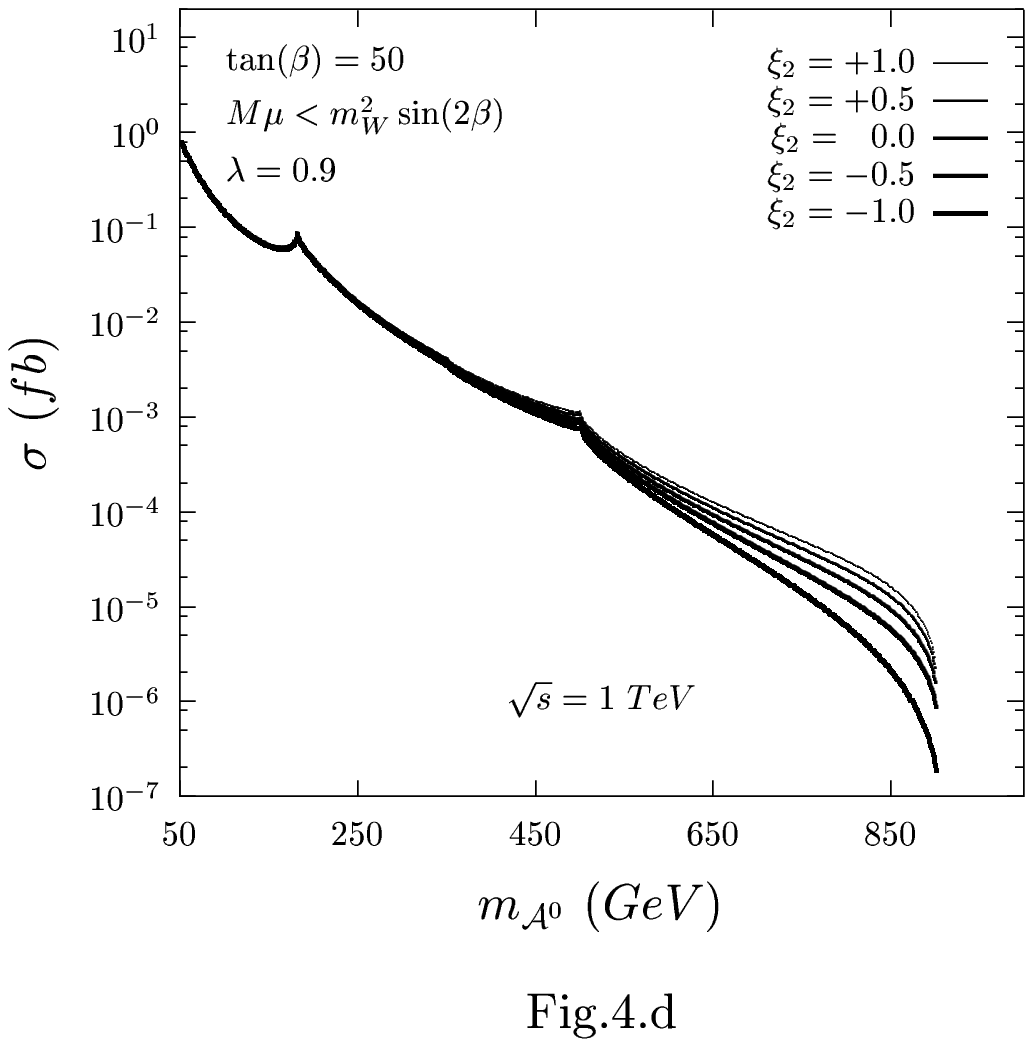}
    \vspace{-2.0cm}
                    
\hspace{4.5cm} Figure 4: Case-2, for $\sqrt{s}=1~TeV$.                                                          

\end{figure}

\newpage

\begin{figure}[thb]

\vspace{20cm}
    \includegraphics{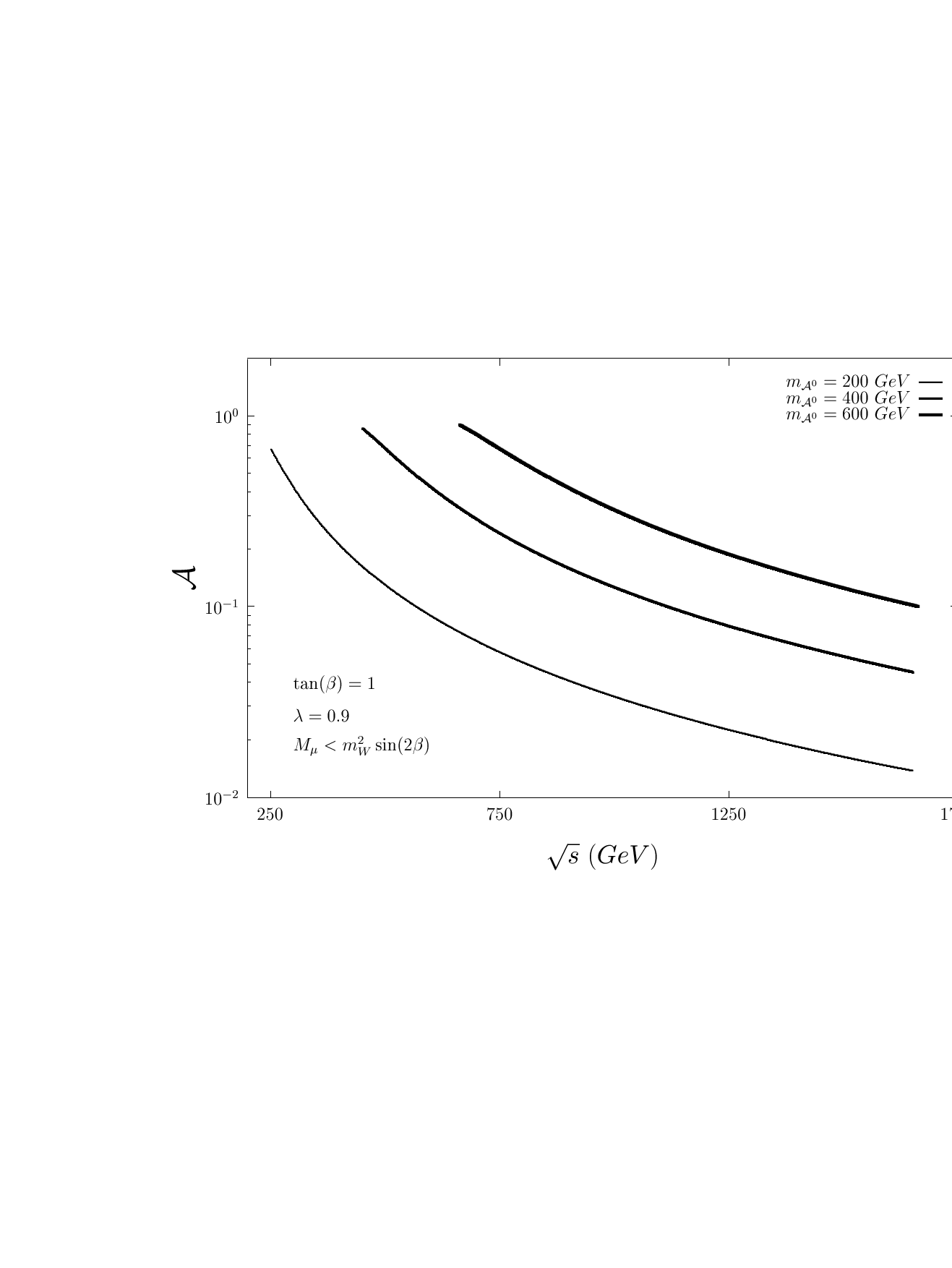}
    \vspace{-9.5cm}

\hspace{5.cm} Figure 5: Polarization asymmetry.

\end{figure}

\newpage
  
\section*{Acknowledgements}
I am deeply indebted to Prof. Dr. T. M. Aliev, for invaluable
contributions and clarifying  discussions 
throughout the whole course of the  work.

\end{document}